\documentclass{optica-article}

\journal{opticajournal} 

\articletype{Research Article}

\usepackage{lineno}

\begin{document}

\title{Flexible experimental platform for dispersion-free temporal characterization of ultrashort pulses}

\author{Patrick Rupprecht,\authormark{1,2,*} Alexander Magunia,\authormark{1} Lennart Aufleger,\authormark{1} Christian Ott,\authormark{1} and Thomas Pfeifer\authormark{1,3}}

\address{\authormark{1}Max-Planck-Institut f{\"u}r Kernphysik, Saupfercheckweg 1, 69117 Heidelberg, Germany\\
\authormark{2}Currently with Lawrence Berkeley National Laboratory, 1 Cyclotron Road, Berkeley, CA 94720, USA\\
\authormark{3}Center for Quantum Dynamics, Ruprecht-Karls-Universit{\"a}t Heidelberg, Im Neuenheimer Feld 226, 69120 Heidelberg, Germany.}

\email{\authormark{*}patrick.rupprecht@mpi-hd.mpg.de} 


\begin{abstract*} 
The precise temporal characterization of laser pulses is crucial for ultrashort applications in biology, chemistry, and physics. 
Especially in femto- and attosecond science, diverse laser pulse sources in different spectral regimes from the visible to the short-wavelength infrared as well as pulse durations ranging from picoseconds to few femtoseconds are employed. 
In this article, we present a versatile temporal-characterization apparatus that can access these different temporal and spectral regions in a dispersion-free manner and without phase-matching constraints. 
The design combines transient-grating and surface third-harmonic-generation frequency-resolved optical gating in one device with optimized alignment capabilities based on a noncollinear geometry.

\end{abstract*}

\section{Introduction}
Ultrashort laser pulses are enabling crucial applications in many different fields of industry and science: from micromachining \cite{gattass2008femtosecond} over multi-photon microscopy \cite{zipfel2003nonlinear} to attosecond physics and chemistry \cite{corkum2007attosecond}.
The temporal characterization of such laser pulses has presented a significant challenge since their first generation in the second half of the 1960s \cite{didomenico1966generation,smith1970mode, french1995generation, diels2006ultrashort}. 
As their duration down to the femtosecond timescale is orders of magnitude shorter than the fastest available electronics, multiple approaches have been developed to address this issue: One class of setups relies on measuring the pulse with itself by means of optical nonlinearities. 
Amongst these techniques are autocorrelation \cite{diels1985control}, frequency-resolved optical gating (FROG) \cite{trebino1993using, trebino1997measuring}, SPIDER \cite{iaconis1998spectral}, 2DSI \cite{birge2006two}, MIIPS \cite{lozovoy2004multiphoton}, ptychographic reconstruction \cite{spangenberg2015ptychographic} and the dispersion-scan approach \cite{miranda2012simultaneous}. 
Further techniques rely on an \textit{in-situ} characterization of the ultrashort pulse in strong-field experiments inside a vacuum environment such as streaking \cite{goulielmakis2004direct} or core-level transient absorption spectroscopy \cite{blattermann2015situ}. 
More recently, tunnel ionization in air has been employed for pulse characterization purposes \cite{park2018direct, cho2019temporal}.
These techniques and their implementations are usually optimized to a specific set of laser parameters like center wavelength, bandwidth, peak intensity, pulse duration, and chirp. 
Nowadays, state-of-the-art table-top setups for ultrafast dynamics experiments rely on multiple laser sources in different wavelength, peak-power and pulse-duration regimes \cite{krausz2009attosecond,calegari2016advances,li2020attosecond}. 
Hence, a common temporal characterization of all available laser-pulse sources imposes a significant challenge to everyday lab operations in modern ultrafast science and especially in attosecond science: In general, the laser front-end is based on the Titanium-Sapphire (Ti:Sa; around 800\,nm center wavelength) \cite{backus1998high} or the Ytterbium (around 1030\,nm) technology \cite{saraceno2019amazing} with primary pulse durations ranging from roughly 20\,fs to 1\,ps. 
These driving pulses are often combined with an optical parametric (chirped) pulse amplification \cite{fattahi2014third} and a nonlinear compression stage \cite{nagy2021high,conejero2018universal} to reach few-cycle pulse durations or even sub-cycle light transients \cite{wirth2011synthesized}. 
Recently, the goal of table-top soft x-ray generation has led to the development of sources in the short-wavelength infrared regime (SWIR; 1\,–\,3\,µm) due to the ponderomotive scaling in the underlying high-order-harmonic generation (HHG) process \cite{shan2001dramatic,popmintchev2010attosecond}. 
As laser pulse sources for attosecond science rely heavily on (extremely) nonlinear optical effects like HHG, temporal pulse-shape characterization and optimization in every stage of the laser system is highly desirable. 
One example includes the HHG of isolated attosecond pulses versus attosecond pulse trains depending on the temporal duration and structure of the driving pulses.
This illustrates the extent of different wavelength and pulse-duration regimes involved in state-of-the-art attosecond science. 
While separate pulse-characterization setups for every component of the respective laser system might be commercially available, a single cost-efficient temporal characterization device which is straight-forward to use and covers all these different pulse sources remained elusive so far. 
In this article, we present such a versatile temporal-characterization apparatus capable of precisely measuring pulses from pico- to few-femtosecond duration and from the visible to the SWIR spectral regime and beyond.

\section{Experimental apparatus}

The pulse-characterization apparatus combines transient-grating (TG) \cite{sweetser1997transient} and surface third-harmonic-generation (sTHG) FROG \cite{tsang1996frequency} in a single device. 
Changing between these two modi is easily conducted via a simple and fast realignment procedure. 
Due to the selection of the sTHG and the TG nonlinear process, no phase-matching constraints apply and hence no fundamental restrictions concerning spectral applicability of the pulse-characterization device exist. 
Furthermore, the apparatus is virtually dispersion-free by relying on reflective optics in combination with a thin optical target. 
This guarantees an accurate measurement of pulses with few- to single-cycle duration. 
The TG approach with its degenerate $\chi^{(3)}$ signal and intuitive traces is suitable for pulses in the visible and near-infrared spectral regions.
Complementary to that, the sTHG nonlinearity as a nondegenerate four-wave-mixing process is ideal for pulse characterization purposes in the SWIR regime since the measured third harmonic is accessible with cost-efficient silicon-based detectors instead of expensive indium-gallium-arsenide (InGaAs) equipped spectrometers.
An illustration of the optical beam path of the versatile pulse-characterization apparatus is presented in Fig.~1. 
\begin{figure}[ht!]
\centering\includegraphics[width=\linewidth]{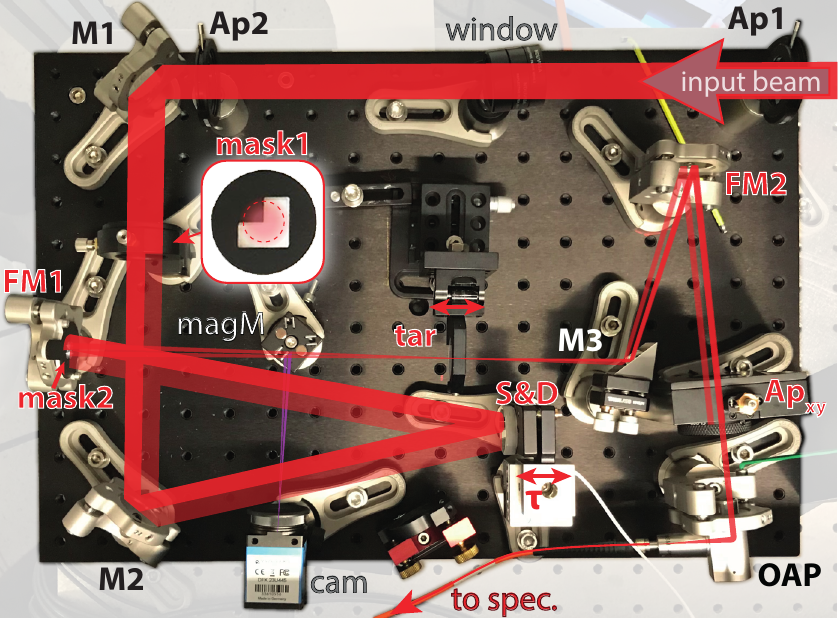}
\caption{Beam path of the pulse-characterization apparatus. The input laser pulses enter from the top right and are aligned on apertures Ap1 and Ap2. To mimic additional dispersive materials in the ultrafast experimental setup and/or to attenuate the beam, windows/ND filters can be inserted in the beam path. Routing mirrors M1 and M2 as well as the one-quadrant mask (mask1; see inset with beam-mode overlay) are realigned to switch between the different pulse-measurement modi. Two \mbox{D-shaped} mirrors (one of them motorized) form the split-and-delay unit (S\&D). The focusing mirror FM1 is equipped with an exchangeable four-hole mask (mask2; see Fig.~2 for details) which splits the beam mode into beamlets. The respective four (alignment mode), three (TG mode) or two (sTHG mode) beamlets are focused into a thin target medium (tar). A magnetic-base mounted mirror (magM) can be inserted to re-route the beams on a camera (cam) to determine the spatial/temporal beam overlap in the target. After the optical target, the beamlets (including the signal beamlet) are routed via mirror M3 and recollimated with the focusing mirror FM2 (the beamlets are shown here for the TG mode). A movable aperture (Ap$_{xy}$) isolates the nonlinear signal and an off-axis parabolic mirror (OAP) focuses the signal into a multimode optical fiber which is connected to a spectrometer.}
\end{figure}
The overall pulse-characterization device is realized on a 30\,cm\,$\times$\,45\,cm breadboard.
In this optical setup, two parts of the input beam mode are time-delayed with respect to each other by two flat \mbox{D-shaped} mirrors.
One of them is moved by a piezoelectric slip-stick actuator (\textit{SmarAct CLS-3232}) with a 21\,mm travel range at a 1\,nm closed-loop sensor resolution. 
For alignment of this split-and-delay unit (S\&D in Fig.~1), a magnetic-based mirror (magM in Fig.~1) re-routes the beam and images the target focus onto a camera (cam in Fig.~1). 
The relative timing of the upper and the lower half of the beam mode are controlled with sub-femtosecond precision over a picosecond range via the spatial displacement of one of the two D-shaped mirrors.
In consequence, the temporal range is limited to 30\,ps only by the spatial walk-off of the foci in the target due to the slightly noncollinear reflective geometry of the S\&D unit. 
The spatial and the temporal overlap of the beamlets at the target position are precisely adjusted by measuring the respective interference pattern with a camera (cam in Fig.~1).
In the next step, the beam mode is spatially split into beamlets with a four-hole mask as shown in Fig.~2(a) and 2(c). 
\begin{figure}[h]
\centering\includegraphics[width=\linewidth]{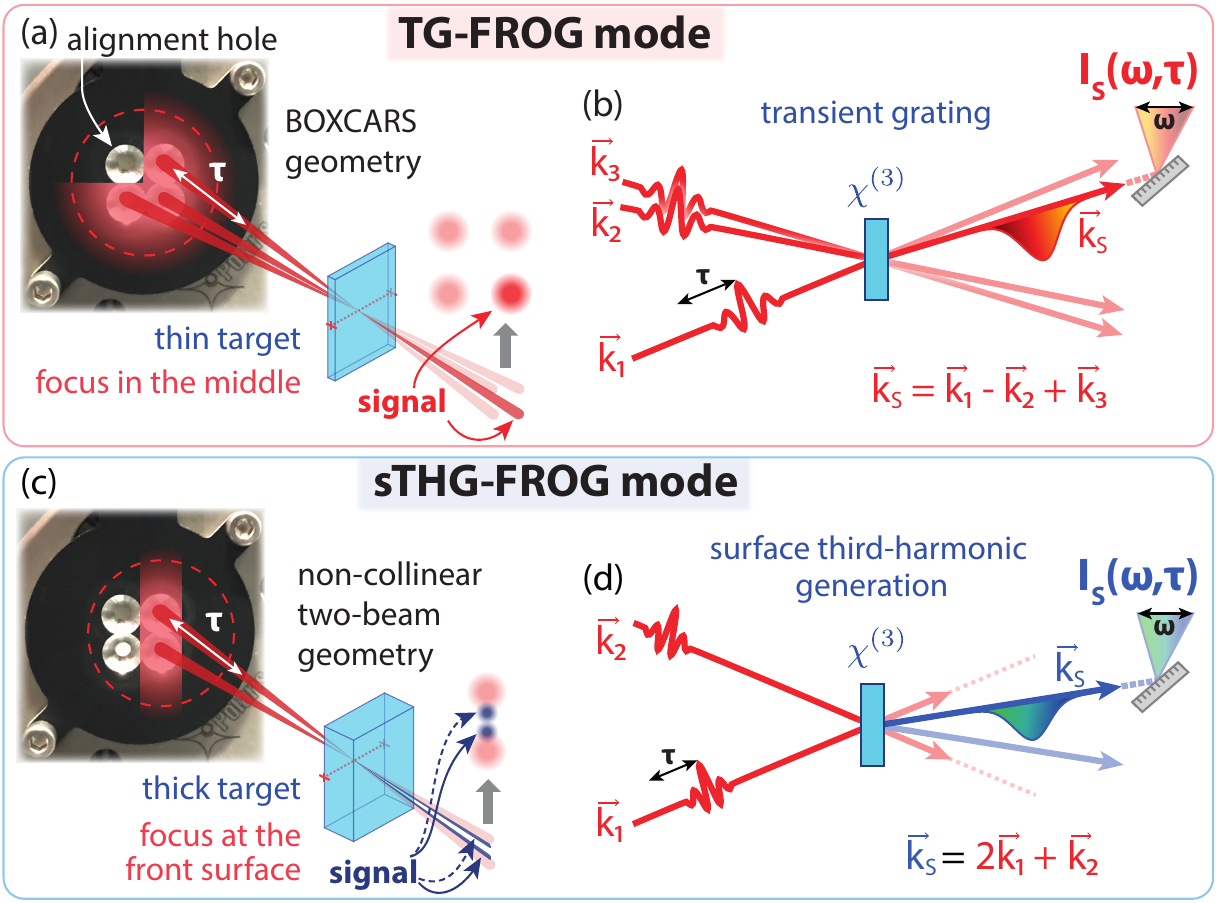}
\caption{Geometry and mechanism of transient-grating and surface third-harmonic-generation mode. Input beam position on the mask (compare \textit{mask2} in Fig.~1) and schematic beam path for (a) the TG and (c) the sTHG mode. Wave-vector schematic, nonlinear-signal generation and spectrally dispersed measurement in (b) TG and (d) sTHG configuration.}
\end{figure}
Depending on the chosen measurement mode, the beam is aligned on the center of the mask (TG mode) or between two vertical holes (sTHG mode). 
This alignment is conducted using two routing mirrors (M1 and M2 in Fig.~1). 
In addition, a one-quadrant mask (mask1 in Fig.~1) is rotated/shifted to ensure the correct beam-mode cutting shown in Fig.~2(a) and 2(c). 
The respective TG or sTHG signals are spatially isolated with an xy-movable, continuously variable iris (Ap$_{xy}$ in Fig.~1).
Thus, switching between sTHG and TG mode is performed by a simple re-alignment procedure within a few minutes without major changes to the pulse-characterization setup.
If the mask1 is completely flipped out of the beam path, the fourth beamlet appears and can be used as pseudo signal for TG-mode alignment all the way to the spectrometer input.
This simplifies the initial alignment of the device significantly.
The four-hole mask is exchangeable to take account of different beam mode sizes.
For an easy and reproducible placement it is directly mounted in front of a focusing mirror (FM1 in Fig.~1). \par
In TG mode, one beamlet is time-delayed with respect to a pair of beamlets that is generating the transient grating in the target medium. 
The noncollinear BOXCARS design \cite{eckbreth1978boxcars} shown in Fig.~2(a) results in the measured dispersion-free transient grating spectrogram \cite{li1999dispersion}

\begin{equation}
	I_{FROG}^{TG}(\omega, \tau) = \left| \int_{-\infty}^{\infty} E(t-\tau)\left| E(t)\right|^2\,e^{-i\omega t} \, dt \right|^2 
	\label{eq:setup_tgFROG}
\end{equation}

\noindent with the angular frequency $\omega$, the time delay between the pulse beamlets $\tau$, and the time-dependent complex electric field of the laser pulse $E(t)$. 
This signal appears at the position of the virtual fourth corner of the BOXCARS-geometry square. 
As the respective TG signal is a degenerate four-wave mixing process, the phase-matching condition is automatically fulfilled and hence an ordinary, thin glass plate can be used as target medium. 
Therefore, no expensive and phase-matching-limited nonlinear crystal is necessary.
Due to the resulting unambiguous and intuitive measurement trace, the TG mode is preferred whenever a suitable spectrometer for the respective fundamental wavelength range is available. \par
In the sTHG mode, the beam realignment results in two noncollinearly focused beamlets [see Fig.~2(c)]. 
By focusing the beamlets on the front surface of a comparably thick target the dispersion-free character of the measurement is maintained while a suppression of the sTHG signal originating from the backside of the target is achieved. 
The sTHG signal is described by

\begin{equation}
	I_{FROG}^{THG}(\omega, \tau) = \left| \int_{-\infty}^{\infty} E(t)^2\,E(t-\tau)\,e^{-i\omega t} \, dt \right|^2 
	\label{eq:setup_thgFROG}
\end{equation}

\noindent and will appear in between the transmitted input beamlets. 
In addition, a weak interferometric surface second-harmonic generation (sSHG) \cite{shen1986surface} signal is present in the middle between the two transmitted input beamlets. 
The sTHG process translates the original spectral domain of the input pulse to 1/3 of its wavelength. 
Thus, this process is ideally suited to characterize SWIR pulses with a conventional silicon-based spectrometer that covers the ultraviolet/visible (UV/VIS) range. 
Photo-detectors based on silicon are limited to wavelengths smaller than $\sim$1100\,nm due to the 1.1\,eV bandgap of silicon. 
The InGaAs-based alternatives that can access the >\,1100\,nm range, however, they are cost intensive and require active cooling. 
While a linear SWIR-related signal cannot be detected by silicon electronics, the respective sTHG signal matches the spectral range of conventional UV/VIS spectrometers. 
In principle, higher-order harmonics (e.g., surface fifth or seventh harmonic) \cite{tsang1996third} can be used to translate even longer mid-IR wavelengths into the UV/VIS regime.

\section{Temporal characterization examples}

The described pulse-characterization setup was built and tested with multiple ultrashort-pulsed laser sources at the \textit{Max Planck Institute for Nuclear Physics} \cite{rupprecht2023wavelength}. 
The respective sources provide amplified laser pulses in the visible to SWIR regime with pulse durations ranging from several picoseconds to the few-femtosecond regime.

\subsection{Ti:Sa pulses from 27\,fs to 8\,ps}

\begin{figure}[ht!]
\centering\includegraphics[width=\linewidth]{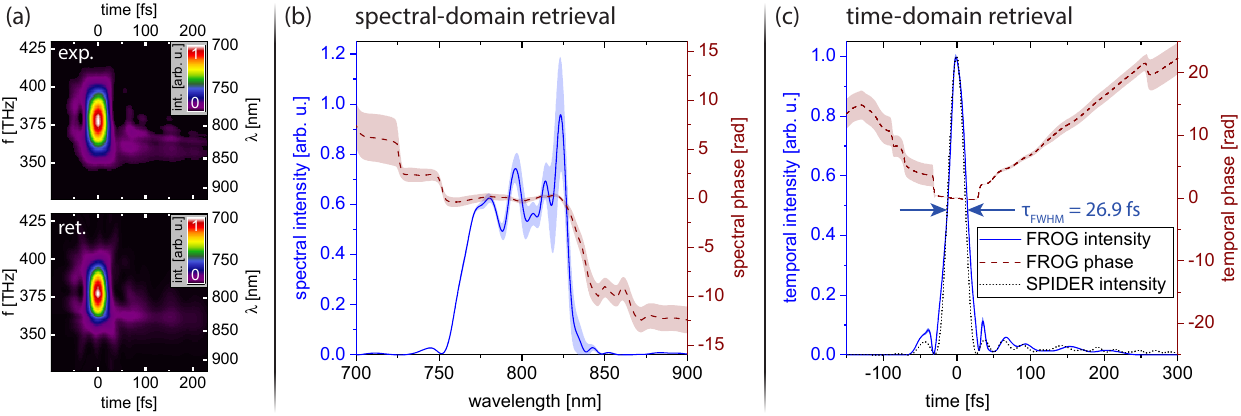}
\caption{TG-FROG measurements of near-Fourier-limited pulses from the Ti:Sa laser system and comparison to a SPIDER measurement. (a) Measured (exp.) and retrieved (ret.) TG-FROG traces. (b) Retrieved intensity and phase in the spectral domain. (c) Retrieved intensity and phase in the time domain. Also, the temporal pulse structure measured with a commercial SPIDER device is shown (black dashed line) for comparison.
\label{fig:800FROG}}
\end{figure}

First, pulses from a commercial 1\,kHz chirped-pulse-amplification Ti:Sa laser system (\textit{Legend Elite Cryo; Coherent Inc.}) were characterized. 
For these pulses centered around 800\,nm, the TG-mode of the pulse-characterization setup was used in combination with a silicon-based, fiber-coupled spectrometer (\textit{Ocean-FX-XR1; Ocean Insight}).
Fig.~\ref{fig:800FROG} shows the measurement and retrieval of the compressed laser output. 
For this measurement, a $\sim100\,\upmu$m-thin microscope cover glass (\textit{CG00C; Thorlabs}) was used as target.
The pulse structure is retrieved using the \textit{common pulse retrieval algorithm} (COPRA) \cite{geib2019common}.
The uncertainty bands in the spectral and temporal retrievals [Fig.~2(b) and 2(c)] are obtained by a statistical analysis of the retrieval results originating from different initial starting pulse guesses for the retrieval algorithm. 
For comparison, these pulses were also characterized with a commercial SPIDER device (\textit{Compact LX SPIDER; APE GmbH}) as shown as dotted black line in Fig.~3(c). 
Here, the temporal pulse structure including pre-/post-pulses from uncorrected higher-order spectral-phase terms matches well for both devices.
The retrieved full-width-at-half-maximum (FWHM) duration of the FROG measurement of $\tau_{FWHM}^{TG}$(800\,nm)~$= (26.9 \pm 0.7)$\,fs is slightly longer than the one extracted from the SPIDER measurement ($\tau_{FWHM}^{SPIDER}$(800\,nm)~=~24.8\,fs).
The FWHM uncertainty of the FROG-deduced duration is given by combining statistical deviations represented by the merely visible blue error band in Fig.~3(c) with a systematic uncertainty of half a time-step size induced by the measurement or the retrieval-algorithm rebinning (whichever value is larger).
A potential source for the small deviations in the retrieved pulse structures resulting from the FROG and the SPIDER measurements is spatio-temporal inhomogenity in the beam mode. 
While the collinear design of the SPIDER device focuses on the center of the beam mode, the noncollinear BOXCARS setup of the FROG apparatus employs off-center regions of the beam mode to generate the measurement signal.\par
In a second step, to elucidate the capabilities of the versatile temporal-characterization platform in the long-pulse limit, heavily chirped pulses from a Ti:Sa laser system were characterized. 
In this case, a thicker target (1\,mm UV fused silica plate), which contributes only a negligible amount of dispersion to extensively chirped pulses, was used.
Apart from that, the same setup configuration was used as for the compressed Ti:Sa measurement presented above (TG-mode; UV/VIS spectrometer).
The measurement result and retrievals are shown in Fig.~4.
\begin{figure}[ht!]
\centering\includegraphics[width=\linewidth]{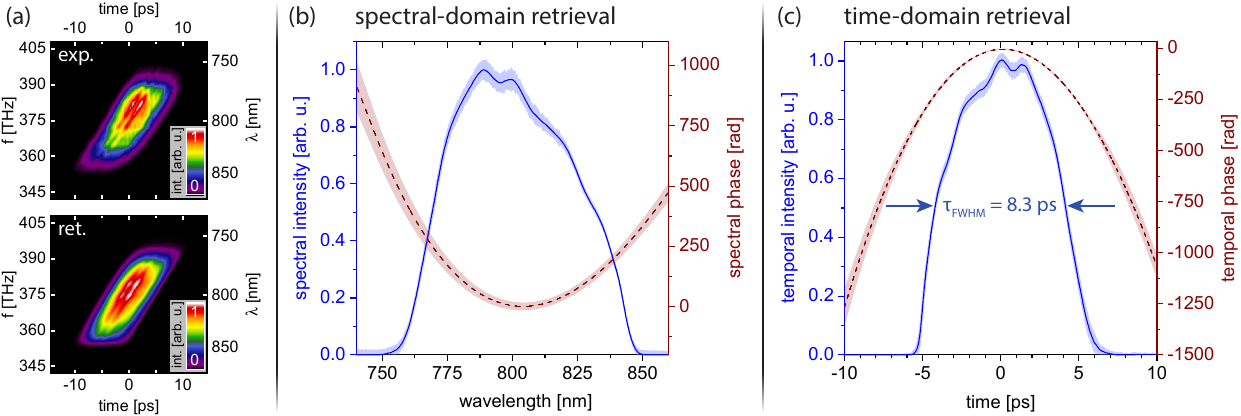}
\caption{TG-FROG measurements of heavily chirped Ti:Sa pulses of picosecond duration. (a) Measured (exp.) and retrieved (ret.) TG-FROG traces. (b) Retrieved intensity and phase in the spectral domain. (c) Retrieved intensity and phase in the time domain.
\label{fig:800FROGchirped}}
\end{figure}
By detuning the grating compressor of the laser system, a group delay dispersion of around GDD~=~$1.5 \times 10^4$\,fs$^2$ and a third-order dispersion of approximately TOD~=~$3.7 \times 10^4$\,fs$^3$ was added to the pulse, resulting in a retrieved pulse duration of $\tau_{FWHM}^{TG}$(800\,nm chirped)~$= (8.3\pm 0.2)$\,ps. 
The slight noncollinear angle of incidence (AOI) of the S\&D unit in Fig.~1 and the associated spatial foci walk-off in the target limits the usable time-delay range to about 30\,ps while the piezoelectric actuator supports up to 140\,ps.
Hence, an optimization for >\,10\,ps FWHM pulse durations by minimizing the mentioned AOI (e.g. by using a longer breadboard) is easily feasible.
Alternatively, a mechanical walk-off compensation via a motorized mirror mount could be employed.

\subsection{Short-wavelength-infrared few-cycle pulses}

In order to showcase the accessibility of different spectral regimes, broadband pulses in the SWIR spectral region were characterized utilizing both, the TG and sTHG mode.
The corresponding pulses are generated by spectrally broadening the output of an optical parametric amplifier via self-phase-modulation in a gas-filled hollow-core capillary and temporally compressing the pulses by propagation through dispersion-compensating materials \cite{rupprecht2023wavelength}.\\
First, a few-cycle pulse with a center wavelength of $\lambda$~=~1550\,nm was characterized using the TG-mode, a $\sim100$\,µm-thin microscopy cover-glass target and an InGaAs spectrometer (\textit{NIRquest 512; Ocean Insight}).
The resulting measurement trace and retrieval is depicted in Fig.~5(a)~-~(c).
\begin{figure}[ht!]
\centering\includegraphics[width=\linewidth]{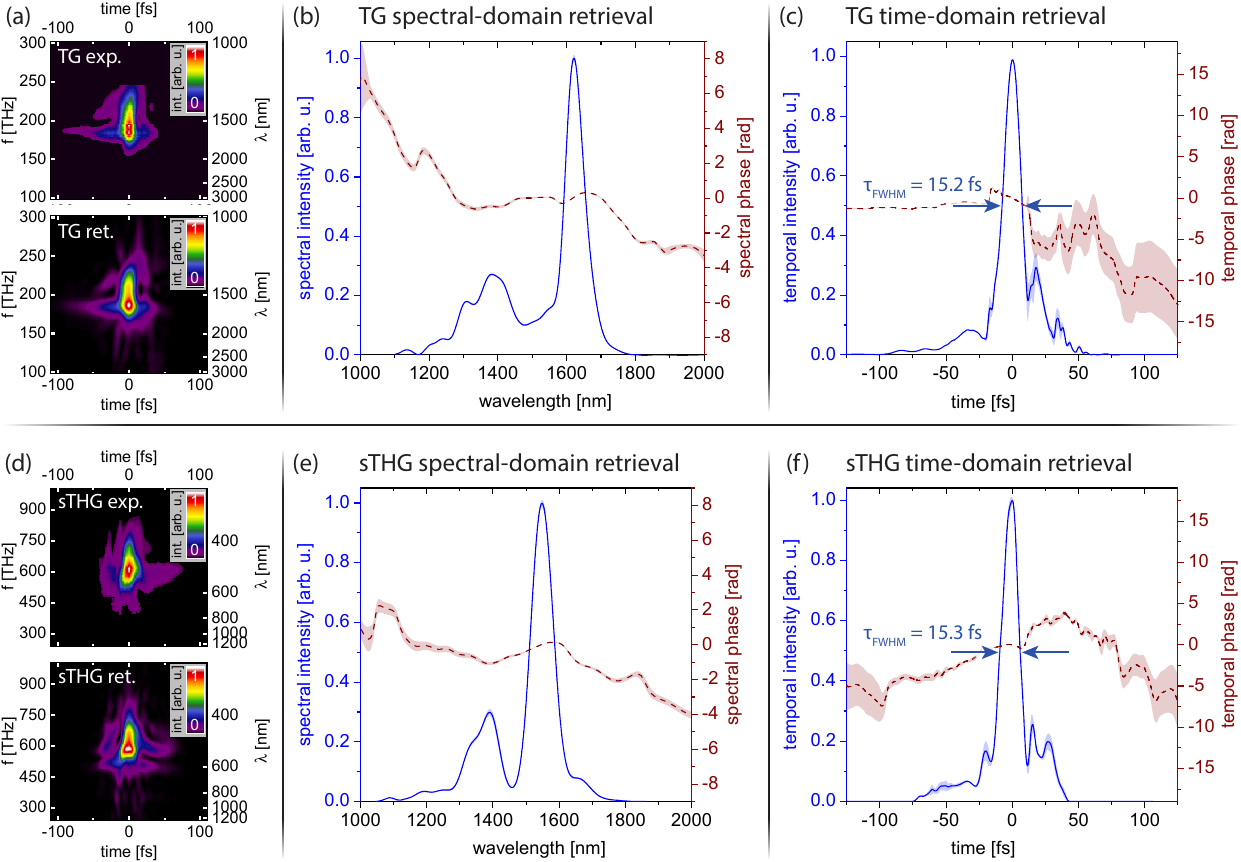}
\caption{TG-FROG versus sTHG-FROG measurement of few-cycle pulses with a center wavelength of 1550\,nm. For the TG measurement, an InGaAs spectrometer was used while the sTHG data were recorded with a silicon-based UV/VIS spectrometer. Measured (exp.) and retrieved (ret.) FROG traces of the TG (a) and sTHG (d) measurements. Retrieved intensity and phase in the spectral domain [TG - (b); sTHG - (e)]. Retrieved intensity and phase in the time domain [TG - (c); sTHG - (f)]. 
\label{fig:1550FROG}}
\end{figure}
Here, the pulses were compressed in the time domain utilizing negative dispersion of fused silica (GVD$_{FS}$(1550\,nm)~=~-28\,fs$^2$/mm). 
By propagation through additional 4\,mm of fused silica, a pulse of three-optical-cycle duration ($\tau_{FWHM}^{TG}$(1550\,nm)~=~$(15.2 \pm 0.5)$\,fs) was achieved. 
The residual TOD, visible in the red-dashed spectral phase in Fig.~5(b), is due to the positive TOD of fused silica even in the SWIR region (TOD$_{FS}$(1550\,nm)~=~151\,fs$^3$/mm).
To compare the different modi of the pulse-measurement apparatus with each other, the same pulses were characterized using the sTHG scheme.
The corresponding results are presented in Fig.~5(d)~-~(f).
Here, a thicker target (2\,mm z-cut MgF$_2$; \textit{United Crystals}) was aligned such that its front surface coincides with the focus position [see Fig.~2(c)].
The same silicon-based UV/VIS spectrometer as for the 800\,nm measurements, described in section 3.1, was used to detect the sTHG signal.
Overall, the TG and the sTHG retrievals match concerning the general temporal pulse structure and the retrieved FWHM duration ($\tau_{FWHM}^{sTHG}$(1550\,nm)~=~$(15.3 \pm 0.5)$\,fs).
The differences in the spectral domain are likely caused by inhomogenities in the beam mode (e.g. spatial chirp) as the sTHG mask cuts different regions of the beam mode compared to the sTHG mask [see Fig.~2(a) and (c)].\par
Due to the limitations of the available InGaAs spectrometer (spectral range from 1.2\,µm to 2.0\,µm), longer-wavelength few-cycle pulses around 2000\,nm center wavelength are only accessible in the sTHG mode of the pulse measurement device. 
Hence, Fig.~6 shows the resulting sTHG measurement trace and retrievals of such an exemplary longer-wavelength few-cycle pulse characterization.
The measured pulse duration of $\tau_{FWHM}^{sTHG}$(2000\,nm)~=~$(15.8 \pm 0.6)$\,fs corresponds to a sub-three-optical-cycles pulse. 
This temporal duration is achieved by propagation through additional 2\,mm of potassium bromide (GVD$_{KBr}$(2000\,nm)~=~38\,fs$^2$/mm). 
Here, the positive GDD introduced by the propagation through KBr is necessary to counteract the negative dispersion by the unavoidable fused silica components in the beam path.
\begin{figure}[ht!]
\centering\includegraphics[width=\linewidth]{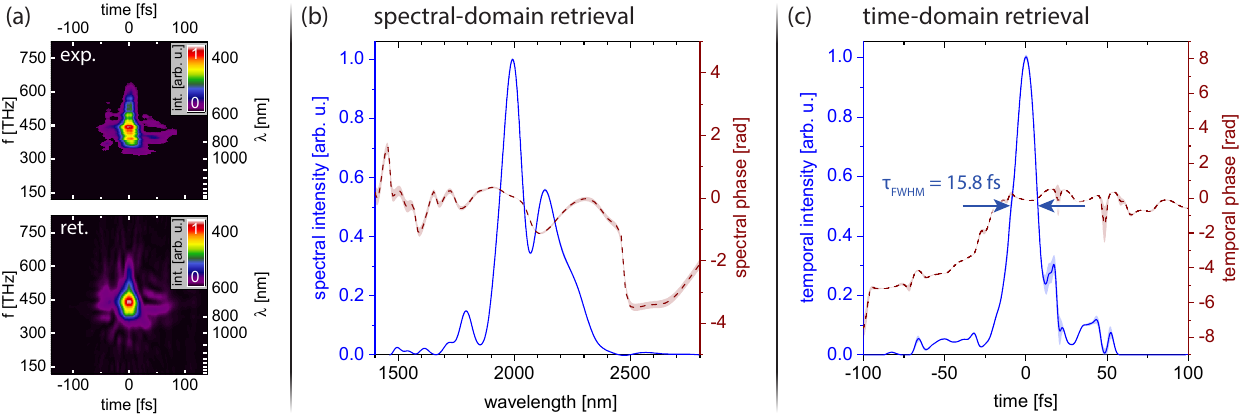}
\caption{Surface THG-FROG measurement of few-cycle pulses with a center wavelength of 2000\,nm. Here, a  silicon-based UV/VIS spectrometer was used for the measurement. (a) Measured (exp.) and retrieved (ret.) TG-FROG traces. (b) Retrieved intensity and phase in the spectral domain. (c) Retrieved intensity and phase in the time domain.
\label{fig:2000FROG}}
\end{figure}

\section{Conclusion}
In this article, we presented a versatile temporal characterization platform which is capable of measuring pulses from the few-femtosecond to the picosecond regime and spanning a broad spectral region from the visible to the SWIR. 
These capabilities were demonstrated by measuring compressed ($\tau_{FWHM}^{TG}$(800\,nm)~$= (26.9 \pm 0.7)$\,fs) as well as chirped ($\tau_{FWHM}^{TG}$(800\,nm~chirped)~$= (8.3\pm 0.2)$\,ps) pulses around 800\,nm from a Ti:Sa laser system in TG-FROG mode by utilizing a silicon-based spectrometer.
Furthermore, few-cycle pulses in the SWIR region were characterized: The TG-mode was applied to a three-cycle ($\tau_{FWHM}^{TG}$(1550\,nm)~=~$(15.2 \pm 0.5)$\,fs) pulse with a center wavelength of 1550\,nm.
Here, the degenerate transient-grating signal was measured with an InGaAs spectrometer and compared with a non-degenerate third-harmonic-generation measurement which utilizes a conventional silicon-based UV/VIS spectrometer ($\tau_{FWHM}^{sTHG}$(1550\,nm)~=~$(15.3 \pm 0.5)$\,fs).
Also, a sub-three-cycle pulse ($\tau_{FWHM}^{sTHG}$(2000\,nm)~=~$(15.8 \pm 0.6)$\,fs) around 2000\,nm center wavelength was characterized using the sTHG mode and a UV/VIS spectrometer.
The presented pulse-characterization platform was crucial for achieving the best possible compression of SWIR few-cycle pulses by balancing their dispersion via propagation through dispersive media such as fused silica (negative GVD for $\lambda < 1270$\,nm) and potassium bromide (positive GVD for $\lambda < 3830$\,nm).
The described pulse-characterization device was successfully employed in the context of table-top time-resolved x-ray absorption spectroscopy experiments: Here, the precise knowledge of the temporal structure of the utilized SWIR pulses contributed to revealing exchange-energy tunability in molecules \cite{rupprecht2022laser} and to disentangling smallest vibrational and electronic signatures in ultrafast molecular dynamics \cite{rupprecht2022resolving}.
This highlights the importance of a universal pulse-characterization device for interpreting the results of state-of-the-art ultrafast experiments at the femto- and attosecond timescales.
Furthermore, due to its transportability, flexibility in operation and the straight-forward alignment, the pulse-characterization setup can be easily operated for various laser systems and laboratories and installed at different positions within an optical setup.\par
As this pulse-characterization platform is not limited by dispersion- or phase-matching bandwidth constraints, its only limitation in regard to minimum and maximum pulse durations as well as spectral applicability is determined by the AOI of the split-and-delay unit, the noncollinear angle of the utilized BOXCARS mask and the spectral reflectivity bandwidth of the mirrors.
Hence, this platform is suitable for the charaterization of pulses down to the deep UV spectral region (e.g., around 200\,nm): Here, pulse energies are typically lower than in the discussed UV/VIS/IR cases making it necessary to split the beam mode efficiently.
This could be accounted for by utilizing a two-hole mask and a self-diffraction scheme \cite{graf2008intense} in combination with UV-enhanced aluminium mirrors. 
As the built-in piezoelectric actuator is vacuum compatible, an in-vacuum use of the device is possible without further modifications, which is crucial for UV pulse characterization purposes.
Moreover, a further miniaturization of the device is possible especially for few-cycle, in-vacuum pulse characterization. 
On the other extreme of the spectral scale, utilizing the sTHG mode with a conventional silicon-based spectrometer grants access to pulses with spectra spanning up to $\sim 3300$\,nm. 
Beyond this wavelength, cost-intensive but readily available InGaAs spectrometers could be used to push the boundary to the 6000\,nm mid-infrared region and beyond. 
An alternative route to extend the measurement capabilities towards the mid-infared is to utilize the surface fifth or seventh harmonic in combination with a silicon-based UV/VIS spectrometer.
\\
\\
\begin{backmatter}
\bmsection{Funding}
We acknowledge financial support by the Deutsche Forschungsgemeinschaft (DFG, German Research Foundation) under Germany’s Excellence Strategy EXC2181/1-390900948 (the Heidelberg STRUCTURES Excellence
Cluster) and by the European Research Council (ERC) (X-MuSiC 616783).
P.~Rupprecht acknowledges funding by the Alexander von Humboldt Foundation (Feodor-Lynen Fellowship).

\bmsection{Acknowledgments}
The authors thank Nils Geib for helpful discussions concerning the COPRA retrieval algorithm.

\bmsection{Disclosures}
The authors declare no conflicts of interest.

\bmsection{Data availability} Data underlying the results presented in this paper are not publicly available at this time but may be obtained from the authors upon reasonable request.

\end{backmatter}


\bibliography{sample}






\end{document}